\documentclass[11pt,a4paper]{article}
\usepackage[hyperref]{acl2020}
\usepackage{times}
\usepackage{latexsym}

\usepackage{microtype}
\usepackage{booktabs}
\usepackage{graphicx}
\usepackage{multirow} 
\usepackage{multicol}
\usepackage{tikz}
\usepackage{latexsym}
\usepackage{amsmath}
\usepackage{amsfonts}
\usepackage{graphicx}
\usepackage{array}
\usepackage{longtable}
\usepackage{subcaption}
\usepackage{bbm}
\usepackage{booktabs} 
\usepackage{multirow}

\newcounter{cdCounter}
\newif\ifcdvar
\cdvartrue
\ifcdvar
\newcommand{\christina}[1]{{\small \color{red} \refstepcounter{cdCounter}\textsf{[CD]$_{\arabic{cdCounter}}$:{#1}}}}
\else
\newcommand{\christina}[1]{}
\fi

\newcounter{fpCounter}
\newif\iffpvar
\fpvartrue
\iffpvar
\newcommand{\fabio}[1]{{\small \color{blue} \refstepcounter{fpCounter}\textsf{[FP]$_{\arabic{fpCounter}}$:{#1}}}}
\else
\newcommand{\fabio}[1]{}
\fi

\newcounter{kpCounter}
\newif\iffpvar
\fpvartrue
\iffpvar
\newcommand{\kashyap}[1]{{\small \color{purple} \refstepcounter{kpCounter}\textsf{[KP]$_{\arabic{kpCounter}}$:{#1}}}}
\else
\newcommand{\kashyap}[1]{}
\fi

\newcounter{lmCounter}
\newif\iffpvar
\fpvartrue
\iffpvar
\newcommand{\louis}[1]{{\small \color{teal} \refstepcounter{lmCounter}\textsf{[LM]$_{\arabic{lmCounter}}$:{#1}}}}
\else
\newcommand{\louis}[1]{}
\fi

\iftrue
\renewcommand{\fabio}[1]{}
\renewcommand{\louis}[1]{}
\renewcommand{\kashyap}[1]{}
\renewcommand{\christina}[1]{}
\fi

\aclfinalcopy 

\newcommand{\system}{\textsc{GET}}

\title{Entity Tagging: Extracting Entities in Text Without Mention Supervision}
\author{Christina Du \\
  Meta AI \\\And
  Kashyap Popat \\
  Meta AI \\\And
  Louis Martin \\
  Meta AI \\\And
  Fabio Petroni \\
  Meta AI \\}

\date{}

\begin{document}
\maketitle
\begin{abstract}

Detection and disambiguation of all entities in text is a crucial task for a wide range of applications.
The typical formulation of the problem (i.e., entity linking) involves two stages: (1) detect mention boundaries, and (2) link all mentions to a knowledge base. For a long time, mention detection has been considered as a necessary step for extracting all entities in a piece of text, even if the information about mention spans is ignored by some downstream applications that merely focus on the set of extracted entities. 
In this paper we show that, in such cases, detection of mention boundaries does not bring any considerable performance gain in extracting entities, and therefore can be skipped.
To conduct our analysis, we propose an ``Entity Tagging" formulation of the problem, where models are evaluated purely on the set of extracted entities without considering mentions. We compare a state-of-the-art mention-aware entity linking solution against \system{}, a mention-agnostic sequence-to-sequence model that simply outputs a list of disambiguated entities given an input context. We find that these models achieve comparable performance when trained both on a fully (i.e., AIDA-YAGO2) and partially annotated dataset (i.e., Wikipedia) across multiple benchmarks, demonstrating that \system{} can extract disambiguated entities with strong performance without explicit mention boundaries supervision. Code, data and pre-trained models are available at \url{https://github.com/facebookresearch/GROOV#get}.
\end{abstract}

\section{Introduction}

\begin{figure*}
\centering
\begin{tikzpicture}[every node/.style={inner sep=6,outer sep=6}]
\node[draw,align=left,text width=5.5cm] at (-16,0) {\textit{input text:}\\
A study published in journal Astronomy \& Astrophysics  last month reported astronomers from the ESO discovered a black hole in the Telescopium constellation. The study stated the black hole is about 1010 $\pm$ 195 light years (310 $\pm$ 60 parsec) away from the Solar System, meaning it is the nearest known black hole from the Earth.};
\node[draw,align=left,text width=9.5cm] at (-7.9,1.15) {\textit{\textbf{Entity Linking} output:}\\ $[21:53]$ Astronomy \& Astrophysics, $[74:85]$ Astronomy, $[95:98]$
 European Southern Observatory, $[112:122]$
Black hole, $[130:155]$ Telescopium, $[178:188]$ Black hole, $[209:220]$ Light-year, $[231:237]$
 Parsec, $[253:265]$ Solar System, $[319:324]$ 
 Earth};
\node[draw,align=left,text width=9.5cm] at (-7.9,-1.65) {\textit{\textbf{Entity Tagging} output:}\\ Astronomy \& Astrophysics, Astronomy, European Southern Observatory, Black hole, Telescopium, Light-year, Parsec, Solar System, Earth};
\end{tikzpicture}
\caption{Example showing the expected output for ``Entity Linking" and ``Entity Tagging" problem. Note that in entity linking the output is a list of entities with mention boundaries, while the output for entity tagging is a set of entities. } 
\label{fig:M1}
\end{figure*}

Extracting all entities in a piece of text is a fundamental building block for several real-world applications, ranging from recommender systems to chat bots, from information retrieval engines to question answering systems ~\citep{ferrucci2012introduction,slawski_2015,yang-etal-2018-collective,chen2017reading,lewis2020retrievalaugmented,roller2020recipes}.
The typical formulation of this problem, also known as ``Entity Linking" (EL), involves (1) detecting all entity mention boundaries, and (2) disambiguating each mention by linking it to a given knowledge base (e.g., Wikipedia) \cite{guo2018robust,wu-etal-2020-scalable,li-etal-2020-efficient,Zhang2021EntQAEL,de-cao-etal-2021-highly}.

Although there exists application scenarios in which mention boundaries are an essential information, such as Wikification ( \citet{mihalcea2007wikify}) or opinion mining, which requires mentions to identify fine-grained opinion polarity towards a specific target \cite{liu2012survey, medhat2014sentiment, zhang2018deep},
other real-world applications often completely ignore mention boundaries and exclusively use the set of extracted entities to gain a high-level understanding of the input \cite{top3}. For instance, semantic search engines such as Google extract relevant entities from input query to understand user intent ~\cite{googleentities}. Similarly, chatbot applications can better understand user needs by identifying the set of entities in conversations ~\cite{kommunicate}.

Despite this distinction, it is generally believed that detecting mentions is a crucial step for extracting all entities from text --- both scenarios are often approached with the same entity linking techniques, which involve detecting mention boundaries, even if that information is ignored in downstream applications~\cite{bunescu2006using,cucerzan2007large,dredze2010entity,hoffart-etal-2011-robust,le2018improving,de-cao-etal-2021-highly}. In our paper, we challenge this common belief by showing that entities can be extracted with performance comparable to state-of-the-art entity linking solutions without considering mention boundaries.



To verify our hypothesis, we propose an ``Entity Tagging" (ET) formulation that evaluates models purely on the set of extracted entities without considering mention boundaries, which better reflects the needs of some real-world applications.
To this end, we create an ET benchmark by re-purposing popular EL datasets to the ET formulation and compare a state-of-the-art mention-aware EL solution \cite{de-cao-etal-2021-highly} against a novel  mention-agnostic sequence-to-sequence model, \system{} (for \textit{Generative Entity Tagging}), that generates 
a set of disambiguated entities mentioned in a given text autoregressively.

Our empirical study shows that \system{} achieves comparable or slightly better performance than the state-of-the-art EL approach in this setting when trained on a fully annotated dataset (i.e., AIDA-YAGO2 \citep{hoffart-etal-2011-robust}) or a partially annotated one (i.e., Wikipedia), indicating that transformer-based sequence-to-sequence models can effectively extract relevant entities without explicit supervision for mention boundaries, potentially opening up a brand new modelling direction for the problem.











In summary, the key contributions of this work are as follows:
\begin{itemize}
    \item We introduce the novel Entity Tagging (ET) formulation and develop a benchmark for evaluating ET systems.
    \item We propose and release \system{}, a mention-agnostic sequence-to-sequence model that can extract the set of unique entities in the input.
    \item We show that the mention-agnostic model performs as well as a state-of-the-art mention-based EL system when trained with fully or partially annotated data.
\end{itemize}

\section{Entity Tagging Task}





We define the ``Entity Tagging" (ET) problem as follows: given a textual input source $x$, a model has to return the set of entities mentioned in $x$ from a collection of entities $\mathcal{E}$. For instance, $\mathcal{E}$ can be the set of Wikipedia articles.  Similarly to \citet{DBLP:conf/iclr/CaoI0P21}, we assume that each $e \in \mathcal{E}$ is uniquely assigned to a textual representation (i.e., its name): a sequence of tokens $y$ (e.g., Wikipedia pages are identified by their titles). 
Note that this formulation differs from the traditional EL problem where models are additionally required to return all mention boundaries for each entity in $x$. For the sake of clarity, Figure \ref{fig:M1} demonstrates an example with the expected output of two formulations.

\section{\system{}: Generative Entity Tagging}

We frame Entity Tagging as a sequence-to-sequence problem and propose \system{}, an autoregressive model that produces a set of entity names given an input sequence.
\system{} is inspired by previous work on generative entity linking \cite{de-cao-etal-2021-highly, DBLP:conf/iclr/CaoI0P21} and autoregressive topic-tagging models \cite{Simig2022}.


\paragraph{Training}
Given input text $X_i$, a set of gold entities $Y_i$ and a permutation $\pi$, the model needs to generate the concatenation of relevant tags in a specific order defined by $\pi$.

For each training example, we uniformly sample a random permutation $\pi$ of the gold entities. The model is trained by maximizing the probability of output sequence, where the set of labels are concatenated in correct ordering. Formally, this method corresponds to a loss function described in the following equations:
\begin{equation}
\begin{aligned}
    \mathcal {L}(\theta)     & =  -\mathop{\mathbb{E}}_{\pi}\left[\log\Bigg( P_{\pi}(Y_i|X_i, \theta) \Bigg)\right] \\
    P_{\pi}(Y_i|X_i, \theta) & = \prod_{k=1}^{|Y_i|} P\Bigl(T[k] \bigg|T[1:k-1], X_i, \theta\Bigr) \\
                           T & = T(Y_i, \pi)
\end{aligned}
\label{fig:expectation_loss}
\end{equation}

Where $T$ is the target output sequence with concatenated entity names after a shuffle defined by permutation $\pi$. $T[k]$ represents k-th token in the output sequence and $\theta$ refers to model parameters.


\paragraph{Inference}
At inference time we decode the model naively by choosing the most likely next token at each decoding step. We then split the produced output text by the separator token, resulting in a set of entity names. To guarantee that each predicted entity is in the given knowledge base, we constrain the generation of each entity name using a prefix tree built on all the entity names in $\mathcal{E}$, similarly to \cite{DBLP:conf/iclr/CaoI0P21}.

\begin{table}[t]
\centering
\resizebox{\linewidth}{!}{   
\begin{tabular}{llc}
\toprule
& \textbf{Dataset} & \textbf{\# Examples} \\
\midrule
\multirow{2}{*}{\textbf{Training}} & AIDA (training split)& 942 \\
& Wikipedia Abstracts & 49,058 \\
\midrule
\multirow{5}{*}{\textbf{Evaluation}} & AIDA (test split) & 230 \\
& WNED-WIKI & 320 \\
& ACE2004 & 36 \\
& AQUAINT & 50 \\
& MSNBC & 20 \\
\bottomrule
\end{tabular}%
}
\caption{Number of instances in training and evaluation datasets in our experiments
}
\label{data_table}
\end{table}

\begin{table*}[]
\centering
\resizebox{\linewidth}{!}{%
\begin{tabular}{llcccccc}
\toprule
\multirow{2}{*}{\textbf{Model}} &
  \multirow{2}{*}{\textbf{Training Data}} &
  \multicolumn{2}{c}{\textbf{in-domain}}  &
  \multicolumn{3}{c}{\textbf{out-of-domain}}  &
  \multirow{2}{*}{\textbf{Avg.}} \\
   &
   &
  \multicolumn{1}{l}{\textbf{AIDA}} &
  \multicolumn{1}{l}{\textbf{WNED-WIKI}} &
  \textbf{ACE2004} &
  \multicolumn{1}{l}{\textbf{AQUAINT}} &
  \multicolumn{1}{l}{\textbf{MSNBC}} &
   \\
  \midrule 
Parallel EL & AIDA  & 50.0 & 12.5 & \textbf{29.4} & 23.2 & 26.0 & \textit{28.2} \\
\system{} & AIDA & \textbf{51.7} & \textbf{14.1} & 25.2 & \textbf{27.8} & \textbf{27.7} & \textbf{\textit{29.3}}  \\
\midrule
Parallel EL & AIDA + Wikipedia & 63.5 & \textbf{36.7} & \textbf{36.2} & 40.8 & 40.1 & \textit{43.5} \\
\system{} & AIDA + Wikipedia & \textbf{65.0} & \textbf{36.7} & 35.7 & \textbf{45.5} & \textbf{49.8} & \textbf{46.5}   \\
\bottomrule
\end{tabular}%
}
\caption{$F_{1}$ scores on in-domain and out-of-domain test datasets for different training settings. The last column shows the macro-averaged $F_{1}$ across all the datasets. We use a beam size of 20 for \system{} and 5 for Parallel EL. 
}
\label{f1_table}
\end{table*}

\section{Experimental Setting}


\paragraph{Training Datasets}

We train our models on two different datasets: (1) the AIDA dataset \cite{hoffart-etal-2011-robust}, which consists of news articles annotated with entities from the YAGO2 knowledge base \cite{hoffart2011yago2} and (2) random selection of $50$k Wikipedia abstracts from the KILT dump ~\cite{petroni-etal-2021-kilt} with partially annotated mentions. Specifically, only the most relevant entities in a Wikipedia article are considered and they are usually only annotated on their first occurrence.
Note that the title of each Wikipedia article is also treated as a gold entity, which has no corresponding mention span in the paragraph. More details about the training data statistics are given in Table~\ref{data_table}.


\paragraph{Evaluation Datasets}
We evaluate the performance on AIDA test set as well as on WNED-WIKI, ACE2004, AQUAINT, and MSNBC datasets introduced by~\citet{guo2018robust}.
These datasets are originally designed for entity linking and provide information about both mention boundaries and entity labels. We adapt them to the entity tagging problem by removing mention boundaries and collecting the entity tags in a set, as illustrated in Figure \ref{fig:M1}. We discard mentions tagged with NIL entity (i.e. not having a corresponding entity in the knowledge base) in these datasets. Table~\ref{data_table} provides more details about the size of the evaluation datasets.


\paragraph{Evaluation Metrics}

We measure the model performance by precision, recall, and $F_{1}$ score between the set of ground truth entities and the predicted set of entities.


\paragraph{\system{} Configuration}
Our GET model is based on pre-trained T5-base architecture \cite{t5paper}. It is trained with a standard cross entropy loss to maximize the likelihood of target sequence. For AIDA dataset, we train the model for 30 epochs, with a batch size of 5 and learning rate of 2e-4. For experiments on Wikipedia dataset, we use a learning rate of 2e-4 and train the model with a batch size of 8 for a maximum of 20 epochs. 

During training, for each instance, we randomly shuffle the gold labels and concatenate the processed set of labels with special separator token as target sequence. Given an input sequence, our model learns to generate a sequence with a concatenation of predicted entities without detecting mention spans.
At inference time, we decode the model using constrained beam search with 5 beams. We retrieve a set of entity tags from output sequence as final prediction. 




\paragraph{Parallel EL Baseline}
We compare the performance of GET with Parallel EL \cite{de-cao-etal-2021-highly}, which is the current state-of-the-art 
entity linking model on AIDA.
Following the traditional entity linking formulation, Parallel EL consists of two modules: (1) a mention detection module based on Longformer encoder and feed-forward network that learns to predict the start position as well as the length of mention spans, and (2) a lightweight LSTM-based entity disambiguation module that generates the entity name autoregressively given mention representation. 

To train the model on the Wikipedia dataset, we use the pre-computed mention table provided by  ~\citet{pershina-etal-2015-personalized} to generate negative samples. 
For mentions that are not included in the mention table, we randomly sample an entity from the global set of candidates as negative sample. We set the learning rate for the Longformer and LSTM in Parallel EL to 1e-4 and 1e-3 respectively. We optimize the parameters by Adam with a weight decay of 1e-2. The model is trained with a batch size of 16 for 25 epochs on Wikipedia. At inference time, we decode the model by constrained beam search without using a mention table to evaluate its ability to generalize on out-of-domain data, where such information is unavailable. 


\paragraph{Knowledge Base}
In our experiments, we use a knowledge base with $470,578$ entities, which is originally proposed in the context of the AIDA dataset \cite{hoffart-etal-2011-robust}.










\section{Results}

We report the $F_{1}$ scores of \system{} and Parallel EL on the evaluation benchmark in Table \ref{f1_table}. We additionally report precision and recall on each dataset to provide more insight into model behavior in Appendix Table \ref{pr_table}. 

\system{} shows comparable performance to Parallel EL across different test datasets when trained only on AIDA dataset, with slightly better results on in-domain data. As highlighted in Table \ref{pr_table}, \system{} achieves much higher precision than Parallel EL on average but lower recall. This is mainly because \system{} predicts fewer number of entities than Parallel EL in general. For instance, the average number of predicted entities on WNED-WIKI is 7 for \system{} and 12 for Parallel EL, respectively. When more training data is available (i.e., AIDA+Wikipedia), performance of both models are improved significantly on all the test datasets and \system{} greatly outperforms the baseline in both in-domain and out-of-domain settings.

These results suggest that it is possible to achieve a comparable or even better performance than a state-of-the-art mention-aware entity linking solution without detecting mention spans.
Moreover, our model is easier to train given it only needs a set of input sequences tagged with entity names, while Parallel EL requires mention table and explicit supervision for mention boundaries during training.

\begin{figure}[h]
\centering
\includegraphics[width=8cm]{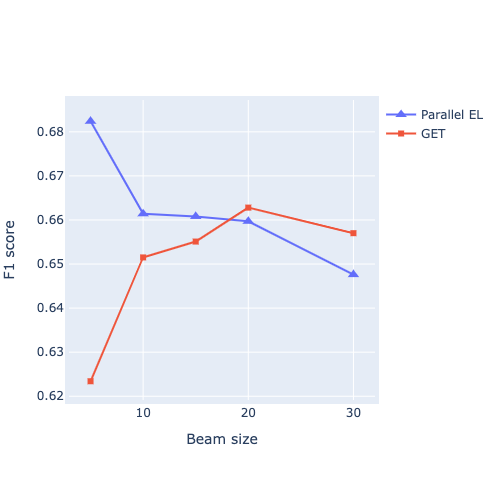}
\caption{$F_{1}$ scores on AIDA validation set against beam size for \system{} and Parallel EL. The best beam size for each model is 20 and 5 respectively. }
\label{fig:ablation}
\end{figure}
\begin{table}[t]
\centering
\resizebox{\linewidth}{!}{   
\begin{tabular}{llc}
\toprule
\textbf{Training Strategy} & \textbf{AIDA Val} & \textbf{AIDA Test} \\
\midrule
Mention Order & 61.6 & 58.7  \\
Random Shuffle & \textbf{66.3} & \textbf{65.0} \\
\bottomrule
\end{tabular}%
}
\caption{
$F_{1}$ scores of \system{} on AIDA with two training strategies: a). sorting target entities based on the natural order of their mentions in text (Mention Order) v.s. b).randomly shuffling the target entity set (Random Order). We use a beam size of 20 and train on AIDA + Wikipedia. 
}
\label{ts_table}

\end{table}

\paragraph{Ablation Study} We investigate the impact of different beam sizes (for both \system{} and Parallel EL) and training strategies (for \system{}) on the performance. 

As shown in Figure \ref{fig:ablation}, the performance of Parallel EL on AIDA validation set gradually declines as the number of beams used for decoding is increased from 5 to 30. In contrast, we observe a steady improvement in the performance of \system{} with the increase of beam size, and the performance tends to stabilize when the beam size becomes larger than 20. Based on these results, we choose the beam size that gives the best scores on AIDA validation data for each model to ensure a fair comparison. Specifically, we decode Parallel EL with 5 beams and use 20 beams for \system{}.

During training, we consider the following strategies to create the output sequence: a).sort the labels according to the order of their mentions b).randomly shuffle the gold labels. Table \ref{ts_table} summarises F1 scores of \system{} model on AIDA dataset in these two scenarios. In this experiment, the model is jointly trained on AIDA and Wikipedia abstracts and decoded with 20 beams. We empirically show that a random permutation of target entities greatly improves the performance.

\section{Related Work}
\paragraph{Entity Linking}
The problem of entity linking has been studied extensively. 
For instance,~\citet{wu-etal-2020-scalable} proposes a bi-encoder based method that encodes the mentions and entities in the semantic space and casts the entity linking problem into the dense retrieval task. Similar approach is followed by~\citet{botha-etal-2020-entity} for multilingual entity linking. On the other hand, more recent approaches solve this problem through autoregressive sequence generation task~\citep{DBLP:conf/iclr/CaoI0P21,de-cao-etal-2022-multilingual}. However, all these approaches assume entity mentions to be available as part of the input.

Very few methods~\citep{kolitsas-etal-2018-end, li-etal-2020-efficient} address the problem of end-to-end entity linking, i.e., having a joint model for mention detection and entity disambiguation. 
More recently,~\citet{de-cao-etal-2021-highly} optimise the generative approach and propose an efficient end-to-end solution for entity linking. However, most of these approaches rely heavily on a mention table (aka alias table) which may not be readily available. 

\paragraph{Extreme Multi Label Classification (XMC)} Extreme Multi Label Classification (XMC) aims to tag content with a subset of labels from an extremely large label set. Given that the target entities in entity tagging are typically from a large knowledge base, e.g., Wikipedia, the problem of entity tagging is very similar to XMC. 

Traditional approaches~\cite{BabSch17,BabSch19,pmlr-v48-yenb16, PPDsparse} for XMC treat each label as a separate class and train a set of binary classifiers for all the labels. These methods largely suffer from increased complexity and large model size. Methods, such as~\citet{parabel,Bonsai,slice}, further reduce the complexity by incorporating various partitioning techniques on the label space. Some of the deep learning-based models~\cite{attentionxml, chang2020taming} have further improved the performance on XMC task. More recently,~\citet{Simig2022} address the problem of incomplete label set and propose an autoregressive model \textsc{groov} for open vocabulary XMC. 

\section{Conclusions}

Although there has been a vast amount of research on entity linking models that can detect mention boundaries when extracting all entities from input text, mention information is sometimes ignored in practical applications.
Even in those cases, it is generally accepted among the research community that detecting mentions boundaries is a crucial step for extracting entities. In this paper we aim to challenge this belief by showing that a mention-agnostic model can extract entities with comparable performance to a state-of-the-art mention-aware solution.
To validate our hypothesis, we introduce a novel task called entity tagging, where models are required to extract a set of entities in the textual input without identifying mention spans and are evaluated purely on the set of entities in output. 
To tackle this task, we propose \system{}, a mention-agnostic entity tagging model that extracts the set of entities from a given text by generating a concatenation of their unique names autoregressively. \system{} shows comparable performance to a cutting-edge mention-aware entity linking model, indicating that large pre-trained language model can effectively extract entities based on the full context without detecting mentions. 
We hope this will lay the foundation for future research on mention-agnostic entity extraction models.

\section*{Limitations}
Our current GET model is incapable of handling long input sequences since the T5 encoder only supports a maximum input length of 512 tokens. However, some test datasets in our benchmark and many real-world applications involve a large number of instances with long context. To adapt our model to real-world scenarios, it is important to explore how to provide support for long input in future work. For example, we can replace T5 encoder in the framework by recent architecture that is able to process long documents \cite{Beltagy2020Longformer}.
%

At inference time, our model is constrained to retrieve entities from a knowledge base of approximately $470$k entries, in line with previous works \cite{hoffart-etal-2011-robust}. However, the coverage of entities only represents a small subset of existing Wikipedia pages, which contains more than $6$M entities for English language. 
Moreover, we focus entirely on a standard monolingual setting (i.e. English entities). How to extend our approach to cross-lingual and multi-lingual environment remains a challenging open problem.


%
%
%

We develop our evaluation benchmark based on existing entity linking datasets, but many publicly available datasets in this field are very small, which can increase the risk of biased evaluation metrics. For example, ACE2004 only contains 36 examples and MSNBC has merely 20 examples. We leave the creation of large-scale benchmarks for entity linking and entity tagging to future work.

We consider a single baseline in our experiments, that is the current state-of-the-art entity linking solution on AIDA \cite{de-cao-etal-2021-highly}. Although more baselines can give additional datapoints in the proposed entity tagging benchmark, the main goal of this paper is not to claim state-of-the-art performance on it but to prove that a mention-agnostic solution can work as well as a strong mention-aware solution in this setting.


\bibliography{anthology,acl2020}
\bibliographystyle{acl_natbib}

\section{Appendix}
\subsection{Experimental Details}
\paragraph{Model}
\system{} model is based on T5-base architecture \cite{t5paper}, which contains about 220M parameters. Parallel EL has a total of 202M parameters, as reported in \cite{de-cao-etal-2021-highly}.

\paragraph{Training Details}
We train the models on AIDA and Wikipedia abstracts with 8 Tesla V100 GPUs. We manually tune the hyperparameters of \system{} based on $F_{1}$ score on AIDA validation set. 

\paragraph{Datasets}
For the experiments on AIDA dataset, we use the public data provided by \cite{de-cao-etal-2021-highly}, which consists of 943 instances for training, 216 for validation and 230 for testing. We create the Wikipedia dataset by randomly sampling about 50k instances from KILT dump \cite{petroni-etal-2021-kilt}. Concretely, the number of instances for training, validation and testing is 49058, 4646 and 4668 after we process the raw data by discarding some data with invalid mention annotations.

\subsection{Results}
Table \ref{pr_table} summarises the precision and recall of \system{} and Parallel EL on different test datasets. We also provide some example output of \system{} and Parallel EL in Table \ref{examples1}, \ref{examples2} and \ref{examples3}.

\begin{table*}[]
\centering
\resizebox{\linewidth}{!}{%
\begin{tabular}{llcccccccccccc}
\toprule
\multirow{2}{*}{\textbf{Model}} &
  \multirow{2}{*}{\textbf{Training Data}} &
  \multicolumn{4}{c}{\textbf{in-domain}}  &
  \multicolumn{6}{c}{\textbf{out-of-domain}}  &
  \multicolumn{2}{c}{\multirow{2}{*}{\textbf{Avg.}}} \\
   &
   &
  \multicolumn{2}{c}{\textbf{AIDA}} &
  \multicolumn{2}{c}{\textbf{WNED-WIKI}} &
  \multicolumn{2}{c}{\textbf{ACE2004}} &
  \multicolumn{2}{c}{\textbf{AQUAINT}} &
  \multicolumn{2}{c}{\textbf{MSNBC}} &
   \\
   & &  P & R & P & R & P & R & P & R & P & R & P & R \\
    \midrule 

Parallel EL & AIDA  & 48.2 & 53.1 & 37.6 & \textbf{31.6} & 23.3 & 46.4 & 40.0 & 30.3 & 28.1 & 37.2 & \textit{35.4} & \textit{39.7} \\
Parallel EL & AIDA + Wikipedia & 58.3 & \textbf{69.7} & 49.8 & 30.9 & 32.1 & \textbf{47.5} & 53.6 & 35.1 & 44.9 & 39.0 & \textit{47.7} & \textbf{\textit{44.4}}\\
\system{} & AIDA  & 59.7 & 50.2 & 22.1 & 11.4 & 24.5 & 30.4 & 41.0 & 22.7 & 36.6 & 25.2 & \textit{36.8} & \textit{28.0}   \\
\system{} & AIDA + Wikipedia & \textbf{74.1} & 61.3 & \textbf{59.7} & 28.5 & \textbf{34.6} & 42.6 & \textbf{64.5} & \textbf{37.1} & \textbf{68.4} & \textbf{41.4} & \textbf{\textit{60.3}} & \textit{42.2} \\
\bottomrule
\end{tabular}%
}
\caption{Precision (P) and recall (R) on in-domain and out-of-domain test datasets for different training settings. The last two columns show the macro-averaged precision and recall score across all the datasets. 
}
\label{pr_table}
\end{table*}

\begin{table*}[]
\centering
\resizebox{\linewidth}{!}
{
\begin{tabular}{r p{0.8\linewidth} }
\toprule
\textbf{Input} &  Prosecutor: Botha may still stand trial. A  prosecutor   on Friday left open the possibility that former apartheid President  P.W. Botha   would be put on trial for abuses committed during  white rule. Prosecutor Jan D'Oliveira said he currently has insufficient evidence to prosecute Botha, but that information collected by the  Truth and Reconciliation Commission (South Africa)  , which investigated apartheid-era abuses, still must be evaluated to ascertain whether charges would ultimately be brought. Last month, the Truth Comission issued a landmark report on its findings. It said Botha had ordered the 1987 bombing of the  African National Congress  ' London headquarters and the 1988 bombing of a  Johannesburg   building housing an anti-apartheid group.  When Botha was head of state in the 1980s, thousands of people were detained without trial. Many were tortured and others killed.  The commission can grant  amnesty   to those who fully confess to politically motivated abuses committed during apartheid. Botha has said he has nothing to confess and has not sought amnesty.  Earlier this year, Botha was convicted of  contempt of court   for refusing to testify before the commission. He was handed a suspended one-year jail sentence and a 10,000  rand  (dlrs 5,700) fine. He has appealed the decision.  (sapa-aos)  \\
\midrule
\textbf{Ground Truth} & 'Amnesty', 'Contempt of court', 'South African rand', 'Johannesburg', 'Prosecutor', 'P. W. Botha', 'Truth and Reconciliation Commission (South Africa)', 'Dominant minority', 'African National Congress' \\
\midrule
\textbf{Parallel EL} & 'African National Congress', 'Johannesburg', 'London', 'P. W. Botha' \\
\midrule
\textbf{\system{}} & 'P. W. Botha', 'Johannesburg', 'South Africa', 'Truth and Reconciliation Commission (South Africa)', 'African National Congress' \\
\bottomrule
\end{tabular}
}
\caption{Predictions of GET and Parallel EL model for a random example in AQUAINT.}
\label{examples2}
\end{table*}
\begin{table*}[]
\centering
\resizebox{\linewidth}{!}
{
\begin{tabular}{r p{0.8\linewidth} }
\toprule
\textbf{Input} & Nembrionic was a Dutch death metal band. They were formed in 1991 under the name Nembrionic Hammerdeath as a black metal band; they changed their name to Nembrionic in 1993. The band released three full-lengths on Displeased Records and one full-length together with Osdorp Posse. They split up in 1999. The band started in April 1991, influenced by Venom, Possessed, and Terrorizer, and recorded a demo in that same year. Their 1992 EP sold 1800 copies and garnered the band a record deal with Dutch metal label Displeased Records. They moved toward grindcore and death metal on Tempter, a split with Consolation. The band played about 30 shows were played to promote the album and went on tour with At the Gates and Consolation. Hammerdeath was dropped from the name in 1995, and their next album, Psycho One Hundred, led to shows at open air festivals such as Liberation Day, Mundial, and Lowlands, and even a show in Ahoy Rotterdam. In 1996, they joined with Osdorp Posse to record Briljant, Hard en Geslepen, which charted in the Netherlands. Nembrionic played at Dynamo Open Air and Pinkpop Festival, and even opened for Slayer in Paradiso, June 1996. Their second to last release was Bloodcult \\
\midrule
\textbf{Ground Truth} & 'Paradiso (Amsterdam)', 'Bloodcult', 'Slayer', 'Death metal', 'Black metal', 'Tempter (album)', 'Briljant, Hard en Geslepen', 'Netherlands', 'Grindcore', 'Possessed (band)', 'Terrorizer', 'A Campingflight to Lowlands Paradise', 'Osdorp Posse', 'Venom (band)', 'Consolation', 'Pinkpop Festival', 'Psycho One Hundred', 'Ahoy Rotterdam', 'Dynamo Open Air', 'Displeased Records', 'At the Gates' \\
\midrule
\textbf{Parallel EL} & 'Netherlands national football team', 'Netherlands', 'Sudan Football Association', 'Pete Sampras', 'Clube Atlético Bragantino', 'Tamil Eelam Liberation Organization', 'Hamburg', 'Confederation of the Greens', 'Revolutionary United Front', 'Olympique de Marseille', 'The Times', 'Ove Olsson' \\
\midrule
\textbf{\system{}} & 'Death metal', 'At the Gates', 'Bloodcult', 'Black metal', 'Displeased Records', 'Attitude (magazine)', 'Nembrionic', 'Terrorizer' \\
\bottomrule
\end{tabular}
}
\caption{Predictions of GET and Parallel EL model for a random example in WNED-WIKI. \louis{We should probably only show the part of the input text and output entities that are interesting to reduce the size.} \fabio{I think it is fine given they are in the appendix}}
\label{examples1}
\end{table*}

\begin{table*}[]
\centering
\resizebox{\linewidth}{!}
{
\begin{tabular}{r p{0.8\linewidth} }
\toprule
\textbf{Input} & British " Euro-sceptic " says Clarke should resign . LONDON 1996-12-06 A " Euro-sceptic " member of the ruling Conservative party said on Thursday British finance minister Kenneth Clarke had to resign to prevent the party disintegrating over the issue of a single European currency . Member of Parliament Tony Marlow said the resignation of the chancellor of the exchequer was the only way to make the Conservatives electable in a general election which must take place by May next year . " We have a divided and split Cabinet . This cannot endure , " Marlow told BBC television 's Newsnight programme on Thursday . " It is not sustainable . Kenneth Clarke has to go . If he does n't resign , the prime minister has got to fire him . " Marlow 's comment come on the heels of speculation that Clarke had threatened to resign if the government changed its " wait and see " policy on a single currency and declared it would not sign up for the currency in the next Parliament . Clarke denied on Thursday he had threatened to resign and said his position on the single currency was in tune with that of Prime Minister John Major . Major told parliament on Thursday he would keep his options open on single-currency membership . His statement was interpreted as a significant victory for Clarke and fellow pro-European Michael Heseltine , deputy prime minister . Pro-European Conservative MP Edwina Currie told the BBC that if Clarke resigned , other ministers would go with him . \\
\midrule
\textbf{Ground Truth} & "John Major", "European Union", "Newsnight", "Cabinet (government)", "United Kingdom", "Kenneth Clarke", "Michael Heseltine", "Antony Marlow", "Edwina Currie", "Euroscepticism", "BBC", "London", "Conservative Party (UK)" \\
\midrule
\textbf{Parallel EL} & "London", "Michael Heseltine", "Celine Dion", "Isaac Hayes", "Richard Marlow", "Isaac Lea Nicholson", "Conservative Party (UK)", "United Kingdom", "John Clarke (actor)", "John Major", "Edwina Currie", "BBC", "Parliament of England", "Anthony Maroon", "Kenneth Clarke", "Alec Eiffel" \\
\midrule
\textbf{\system{}} & "Edwina Currie", "Chancellor of the Exchequer", "Michael Heseltine", "Euroscepticism", "United Kingdom", "London", "Eurozone", "John Major", "Conservative Party (UK)", "BBC", "Kenneth Clarke" \\
\bottomrule
\end{tabular}
}
\caption{Predictions of GET and Parallel EL model for a random example in AIDA test set.}
\label{examples3}
\end{table*}

\end{document}